\documentclass[twocolumn,showpacs,preprintnumbers,amsmath,amssymb]{revtex4}

\usepackage{graphicx}
\usepackage{dcolumn}
\usepackage{bm}

\newtheorem{theorem}{Theorem}

\begin{document}


\title{Multiple-Access Bosonic Communications}

\author{Brent J. Yen}
 \affiliation{
Princeton University\\
Department of Electrical Engineering\\
Princeton, New Jersey 08544}

\author{Jeffrey H. Shapiro}
\affiliation{ Massachusetts Institute of Technology, Research Laboratory of Electronics\\
Cambridge, Massachusetts 02139}

\date{\today}

\begin{abstract}
The maximum rates for reliably transmitting classical information
over Bosonic multiple-access channels (MACs) are derived when the
transmitters are restricted to coherent-state encodings.  Inner
and outer bounds for the ultimate capacity region of the Bosonic
MAC are also presented.  It is shown that the sum-rate upper bound
is achievable with a coherent-state encoding and that the entire
region is asymptotically achievable in the limit of large mean
input photon numbers.
\end{abstract}

\maketitle

\section{Introduction}

The lossy Bosonic channel provides a quantum model for optical
communication systems that rely on fiber or free-space
propagation.  For the pure-loss case, in which the quantum noise
accompanying the loss is the minimum permitted by quantum
mechanics, the classical information-carrying capacity of this
channel has been derived, and shown to be achievable with
single-use coherent-state encoding \cite{lossychannel}.  For the
more general thermal-noise channel, in which the environment
injects an isotropic Gaussian noise, the Holevo information of
single-use coherent-state encoding is a lower bound on the channel
capacity that is tight in the limits of low and high noise levels
\cite{freespace,qcmc04}.  Moreover, if a recent conjecture
concerning the thermal-noise channel's minimum output entropy is
correct, then single-use coherent-state encoding is capacity
achieving \cite{outputentropy}.

To date there has been almost nothing reported about the classical
information-carrying capacity region of multiple-access Bosonic
channels, i.e., Bosonic channels in which two or more senders
communicate to a common receiver over a shared propagation medium.
In this paper we derive single-mode and wideband capacity results
for such channels \cite{byen}. First, we show that single-use
coherent-state encoding with joint measurements over entire
codewords achieves the sum capacity and provides lower bounds on
the individual-user capacities.  Then we quantify the capacity
region that is lost when heterodyne or homodyne detection is
employed---in lieu of the optimum joint measurement---with
single-use coherent-state encoding.   Finally, we derive upper
bounds on the individual-user capacities, and show that they can
be achieved---in the limit of high input photon numbers---by means
of squeezed-state encoding and homodyne detection.

\section{Coherent-State MAC}
\label{sec:coherentstatemac}

\begin{figure}
\centering
\includegraphics[width=6cm]{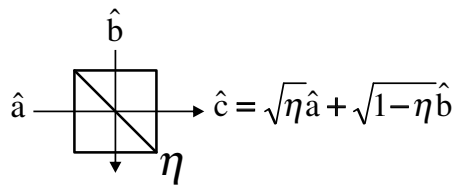}
\caption{\small Two-user, single-mode, optical multiple-access
channel.  Transmitters Alice and Bob have access to input modes
$\hat{a}$ and $\hat{b}$, respectively.  Charlie receives the
output mode $\hat{c} = \sqrt{\eta}\,\hat{a} +
\sqrt{1-\eta}\,\hat{b}$.} \label{mac}
\end{figure}

We will begin with the single-mode optical MAC, shown in
Fig.~\ref{mac}, in which two senders, Alice and Bob, transmit
classical information to a common receiver, Charlie, by accessing
different input ports of a lossless beam splitter with
transmissivity $\eta$, where $0\leq \eta\leq 1$.  The input-output
relation for the electromagnetic modes associated with this
channel is $\hat{c} = \sqrt{\eta}\,\hat{a} +
\sqrt{1-\eta}\,\hat{b}$, where $\hat{a}$ and $\hat{b}$ are the
annihilation operators of Alice's and Bob's input modes, and
$\hat{c}$ is the annihilation operator of the mode that Charlie
measures.
In this section, we derive the capacity of the optical MAC when
Alice and Bob encode complex-valued input messages $\alpha$ and
$\beta$ as coherent states
$|\alpha\rangle_A\otimes|\beta\rangle_B$ with independent
probability densities $p_A(\alpha)$ and $p_B(\beta)$. This
encoding puts the $\hat{c}$ mode in the coherent state
$|\sqrt{\eta}\,\alpha + \sqrt{1-\eta}\,\beta\rangle_C$, so we will
refer to this system as the two-user, single-mode, coherent-state
MAC.

\subsection{Quantum MAC Capacity Theorem}

The capacity region of a two-user multiple-access channel is
defined to be the closure of all rate pairs $(R_1, R_2)$ for which
arbitrarily small error probabilities are achievable in the limit
of long codewords \cite{coverthomas}.  Winter's quantum MAC
capacity theorem \cite{winter} gives the capacity region of a
quantum MAC optimized over arbitrary receiver measurements, and
over codewords that are not entangled over multiple channel uses.
Winter's result presumes a finite-dimensional state space, whereas
the Bosonic MAC has an infinite-dimensional state space.
Nevertheless, we shall rely on his result, which can be extended
to Bosonic channels by means of a limiting argument, see Appendix.
Thus, the capacity region of the two-user, single-mode, optical
MAC from Fig.~\ref{mac} will be taken to be the convex closure of
all rate pairs $(R_1, R_2)$ that satisfy the following
inequalities:
\begin{subequations}
\label{quantummac}
\begin{equation}
   R_1 \leq \int \!p_B(\beta) S(\hat{\rho}_{\beta}^B) \,{\rm d}\beta
   - \int\!\!\!\int \!p_A(\alpha)p_B(\beta)
                      S(\hat{\rho}(\alpha,\beta)) \,{\rm d}\alpha\, {\rm d}\beta
    ,
\end{equation}
\begin{equation}
   R_2 \leq \int\!p_A(\alpha) S(\hat{\rho}_{\alpha}^A) \,{\rm d}\alpha
   - \int\!\!\!\int \!p_A(\alpha)p_B(\beta)
                      S(\hat{\rho}(\alpha,\beta)) \,{\rm d}\alpha\, {\rm d}\beta
   ,
\end{equation}
\begin{equation}
   R_1 + R_2 \leq S(\bar{\rho})
   - \int\!\!\!\int \!p_A(\alpha)p_B(\beta)
                      S(\hat{\rho}(\alpha,\beta)) \,{\rm d}\alpha\, {\rm d}\beta
   ,
\end{equation}
\end{subequations}
for some product distribution, $p_A(\alpha)p_B(\beta)$, on Alice's
and Bob's complex-valued inputs. In these expressions, $S(\cdot)$
is the von Neumann entropy, and the average density operators are
\begin{align}
   \hat{\rho}_{\beta}^B &= \int\! p_A(\alpha) \hat{\rho}(\alpha,\beta) \,{\rm d}\alpha,
   \label{condB}\\
   \hat{\rho}_{\alpha}^A &= \int\! p_B(\beta) \hat{\rho}(\alpha,\beta) \,{\rm d}\beta,
   \label{condA}\\
   \bar{\rho} &= \int\!\!\int \!p_A(\alpha)p_B(\beta)
                      \hat{\rho}(\alpha,\beta) \,{\rm d}\alpha\, {\rm d}\beta,
                      \label{averageAB}
\end{align}
where $\hat{\rho}(\alpha,\beta)$ is the received state given that
messages $\alpha$ and $\beta$ have been transmitted.  The capacity
region will diverge unless the inputs are constrained, so here we
assume that Alice and Bob are subject to the average photon-number
constraints $\langle\hat{a}^\dagger\hat{a}\rangle \le \bar{n}_A$
and $\langle\hat{b}^\dagger\hat{b}\rangle \le \bar{n}_B$,
respectively.

Equations~\eqref{quantummac} constitute the multiple-access
version of the Holevo-Schumacher-Westmoreland theorem, which gives
the classical capacity of a single-user quantum channel
\cite{HSW1,HSW2,HSW3}.  For the coherent-state MAC,
$\hat{\rho}(\alpha,\beta) = |\sqrt{\eta}\,\alpha +
\sqrt{1-\eta}\,\beta\rangle\langle\sqrt{\eta}\,\alpha +
\sqrt{1-\eta}\,\beta|$ is a pure state, so that the second terms
on the right-hand sides of these equations vanish, and the average
density operators in the first terms are found by performing the
indicated integrations in Eqs.~(\ref{condB})--(\ref{averageAB}).

\subsection{Coherent-State MAC Capacity}
\label{sec:singlemode}

Suppose that Charlie uses homodyne or heterodyne detection.  These
are single-use measurements that may not achieve the capacity
region of the coherent-state MAC, but they are easily realized
with existing technology and their capacity regions are simple to
derive.  In particular, coherent-state MACs that use homodyne or
heterodyne detection reduce to classical additive Gaussian noise
MACs:  a scalar Gaussian MAC, with noise variance 1/4, for
homodyne detection, and a 2D white-Gaussian noise MAC, with noise
variance 1/2 per dimension, for heterodyne detection.  It follows
that the capacity region for the coherent-state MAC with homodyne
detection is the set of rate pairs $(R_1, R_2)$ that satisfy
\cite{coverthomas}
\begin{subequations}
\begin{align}
   R_1 &\leq \frac{1}{2} \log(1 + 4\eta \bar{n}_A)\\
   R_2 &\leq \frac{1}{2} \log(1 + 4(1-\eta) \bar{n}_B)\\
   R_1 + R_2 &\leq \frac{1}{2} \log(1 + 4\eta \bar{n}_A +
4(1-\eta)\bar{n}_B),
\end{align}
\end{subequations}
and the capacity region for the coherent-state MAC with heterodyne
detection is
\begin{subequations}
\begin{align}
   R_1 &\leq \log(1 + \eta \bar{n}_A) \\
   R_2 &\leq \log(1 + (1-\eta) \bar{n}_B) \\
   R_1 + R_2 &\leq \log(1 + \eta \bar{n}_A + (1-\eta) \bar{n}_B),
   \label{hetref}
\end{align}
\end{subequations}
when Alice and Bob are subject to the average photon-number
constraints $\bar{n}_A$ and $\bar{n}_B$, respectively.

The preceding two-user results extend easily to the $m$-user
coherent-state MAC that employs homodyne or heterodyne detection.
Here, the $i$th transmitter sends coherent state
$|\alpha_i\rangle$, for $1\le i \le m$, resulting in the channel's
output mode being in the coherent state $|\sum_{i=1}^m
\sqrt{\eta_i}\,\alpha_i\rangle$, where the transmissivities
$\{\eta_i\}$ sum to one.  The $m$-user capacity region with
homodyne detection is then the set of rates $(R_1,\ldots, R_m)$
that satisfy the inequalities
\begin{equation}
    \sum_{i\in S} R_i \leq \frac{1}{2}
                            \log\!\left(
                             1 + 4\sum_{i\in S}\eta_i \bar{n}_i
                             \right),
\end{equation}
for all subsets $S\subseteq \{1,\ldots,m\}$, where $\langle
\hat{a}_i^\dagger\hat{a}_i\rangle \le \bar{n}_i$ is the average
photon-number constraint on the $i$th user.  Similarly, the
capacity region with heterodyne detection is given by the
inequalities
\begin{equation}
    \sum_{i\in S} R_i \leq
                            \log\!\left(
                             1 + \sum_{i\in S}\eta_i \bar{n}_i
                             \right),
\end{equation}
for all subsets $S\subseteq \{1,\ldots,m\}$.

\begin{figure}
\centering
\includegraphics[width=7cm]{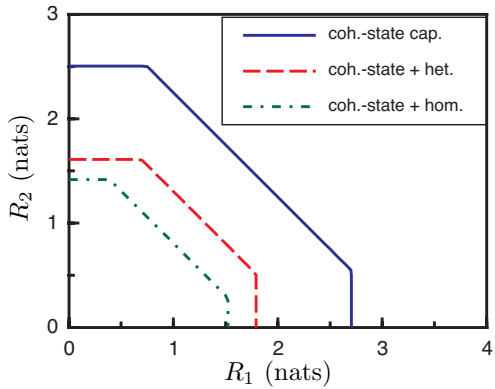}
\caption{\small (Color online) Coherent-state capacity region for
the optical MAC.  The capacity region with optimum reception
(solid line) is given by inequalities \eqref{coherentstatemac}.
The capacity regions with homodyne detection and heterodyne
detection are also shown. This figure assumes $\eta = 1/2$,
$\bar{n}_A = 10$, and $\bar{n}_B = 8$.  Rates are measured in
nats, i.e., logarithms are taken base $e$.} \label{opticalmac1}
\end{figure}

The homodyne and heterodyne detection capacity regions for the
coherent-state MAC provide inner bounds on that channel's capacity
region if no constraints are placed on the receiver measurements.
We will now find the capacity region for the coherent-state
MAC---without restricting the choice of receiver structure---from
the previously stated capacity theorem.  The channel outputs of
the coherent-state MAC are the pure states
$\hat{\rho}(\alpha,\beta) = |\sqrt{\eta}\,\alpha +
\sqrt{1-\eta}\,\beta\rangle\langle\sqrt{\eta}\,\alpha +
\sqrt{1-\eta}\,\beta|$.
It is then easy to show that the circularly-symmetric Gaussian
distributions
\begin{align}
  p_A(\alpha) &= \frac{1}{\pi \bar{n}_A} \exp\!\left(-\frac{|\alpha|^2}{\bar{n}_A}\right), \label{inputa} \\
  p_B(\beta)  &= \frac{1}{\pi \bar{n}_B} \exp\!\left(-\frac{|\beta|^2}{\bar{n}_B}\right), \label{inputb}
\end{align}
are the optimal input distributions, i.e., they maximize the
right--hand sides of \eqref{quantummac} for the coherent-state
MAC. Using these input distributions we find that the capacity
region of the coherent-state MAC, with optimal (joint measurements
over entire codewords) reception is the set of all rate pairs
satisfying
\begin{subequations}
\label{coherentstatemac}
\begin{align}
  R_1 &\leq g(\eta \bar{n}_A) \\
  R_2 &\leq g((1-\eta) \bar{n}_B) \\
  R_1 + R_2 &\leq g(\eta \bar{n}_A + (1-\eta)\bar{n}_B),
\label{coherentstatemac12}
\end{align}
\end{subequations}
where $g(x) \equiv (x+1)\log(x+1) - x\log(x)$ is the Shannon
entropy of the Bose-Einstein probability distribution.
Figure~\ref{opticalmac1} compares the capacity regions for the
coherent-state MAC when homodyne detection, heterodyne detection,
and optimal reception are used.

Users transmitting information over the optical MAC must each
contend with interference created by the other users who are
attempting to access the channel. This type of noise, called
multiple-access interference, is responsible for the pentagonal
shape of the capacity region seen in Fig.~\ref{opticalmac1}.  In
general, users communicating over a multiple-access channel will
encounter channel noise in addition to multiple-access
interference.  A two-user, single-mode, optical MAC that
introduces additional white-Gaussian noise can be modeled by the
evolution equation $\hat{c} = \sqrt{\eta}\;\hat{a} +
\sqrt{1-\eta}\;\hat{b} + \xi$, where $\xi$ is additive classical
zero-mean, complex-valued, white-Gaussian noise with variance
$\langle|\xi|^2\rangle = N$.  Our derivation of the two-user
capacity region for the coherent-state MAC generalizes to include
the presence of additive white-Gaussian noise, with the following
result for the capacity region:
\begin{subequations}
\begin{align}
   R_1 &\leq g(\eta \bar{n}_A + N) - g(N)\\
   R_2 &\leq g((1-\eta)\bar{n}_B + N) - g(N)\\
   R_1 + R_2 &\leq g(\eta\bar{n}_N + (1-\eta)\bar{n}_B + N) -
   g(N).
\end{align}
\end{subequations}

\subsection{Wideband Capacity}

Now let us turn our attention to the wideband coherent-state MAC,
in which Alice and Bob may employ photons of any frequency,
subject to constraints, $P_A$ and $P_B$, on their average
transmitted powers. For a frequency-multiplexed scheme, in which
the radian-frequency domain is divided into bins of width $\Delta
= 2\pi/T$, the channel output for the $i$th mode is
\begin{equation}
   \hat{c}_i = \sqrt{\eta}\;\hat{a}_i + \sqrt{1-\eta}\;\hat{b}_i,
\end{equation}
where $\hat{a}_i$ and $\hat{b}_i$ are the input modes at frequency
$\omega_i = i2\pi/T$, for $i = 1,2,3,\ldots,$ and the
transmissivity $\eta$ is frequency independent.  The average power
constraints on Alice and Bob are given by
\begin{align}
  \sum_i  \hbar\omega_i \langle|\alpha_i|^2\rangle\Delta/2\pi &\leq P_A  \label{powerA}\\
  \sum_i  \hbar\omega_i \langle|\beta_i|^2\rangle\Delta/2\pi &\leq P_B,
\end{align}
where Alice and Bob allocate average photon numbers
$\bar{n}_A(\omega_i) = \langle|\alpha_i|^2\rangle$ and
$\bar{n}_B(\omega_i) = \langle|\beta_i|^2\rangle$, respectively,
to frequency-bin $\omega_i$.

We first derive the capacity region of the wideband coherent-state
MAC with homodyne detection.  When homodyne detection is employed,
the wideband coherent-state MAC is equivalent to a set of parallel
classical MACs with independent zero-mean Gaussian noise, for
which we have derived upper bounds on the individual rates, $R_1$
and $R_2$, and on the sum rate, $R_1 + R_2$, from separate
Lagrange multiplier calculations. In the limit $\Delta \rightarrow
0$, these upper bounds become
\begin{subequations}
\label{widehom}
\begin{align}
  R_1  &\le \sqrt{\frac{\eta P_A}{\pi\hbar}}\\
  R_2  &\le \sqrt{\frac{(1-\eta) P_B}{\pi\hbar}} \\
  R_1 + R_2 &\le \sqrt{\frac{\eta P_A + (1-\eta)P_B}{\pi\hbar}},
\end{align}
\end{subequations} with
\begin{equation}
\eta \bar{n}_A(\omega) = \frac{1}{\omega}\sqrt{\frac{\pi\eta
P_A}{\hbar}} - \frac{1}{4}, \label{widebandwater}
\end{equation}
for $\omega \leq 4\sqrt{\pi\eta P_A/\hbar}$,
\begin{equation}
 (1-\eta) \bar{n}_B(\omega) = \frac{1}{\omega}\sqrt{\frac{\pi(1-\eta)P_B}{\hbar}} - \frac{1}{4},
\end{equation}
for $\omega \leq 4\sqrt{\pi(1-\eta)P_B/\hbar}$, and
\begin{equation}
  \bar{n}'_{AB}(\omega) = \frac{1}{\omega}\sqrt{\frac{\pi[\eta P_A + (1-\eta)P_B]}{\hbar}} -
  \frac{1}{4},
             \label{widebandwater12}
\end{equation}
for $\omega \leq 4\sqrt{\pi[\eta P_A + (1-\eta)P_B]/\hbar}$, where
$\bar{n}'_{AB}(\omega) \equiv \eta\bar{n}_A(\omega) +
(1-\eta)\bar{n}_B(\omega)$.

We see that the optimal mean photon number allocations,
$\bar{n}_A(\omega)$ and $\bar{n}_B(\omega)$, are given by
water-filling formulas, as is found in classical information
theory. The rates \eqref{widehom} define a pentagonal region which
serves as an outer bound for the capacity of the wideband
coherent-state MAC with homodyne detection. To prove that this
outer bound is, in fact, the capacity region, we must show that
Eqs.~(\ref{widebandwater})--(\ref{widebandwater12}) can be
satisfied simultaneously. The average photon number allocations
$(\bar{n}_A(\omega), (\bar{n}'_{AB}(\omega) -
\eta\bar{n}_A(\omega))/(1-\eta))$ for Alice and Bob, achieve the
lower-right corner point
\begin{equation}
   \left(\sqrt{\frac{\eta P_A}{\pi\hbar}},\;
         \sqrt{\frac{\eta P_A + (1-\eta)P_B}{\pi\hbar}} - \sqrt{\frac{(1-\eta)P_B}{\pi\hbar}}
   \right)
\end{equation}
of the outer bound. Similarly, $((\bar{n}'_{AB}(\omega) -
(1-\eta)\bar{n}_B(\omega))/\eta, \bar{n}_B(\omega))$ achieves the
upper-left corner. Thus, the entire region is achievable and hence
is equal to the capacity region. A similar derivation shows that
the wideband coherent-state MAC with heterodyne detection has a
capacity region that is identical to that of homodyne detection.

The preceding two-user wideband results readily extend to the
$m$-user wideband coherent-state MAC.  Suppose that the $k$th user
sends coherent states $\{|\alpha_{k,i}\rangle\}$ across the
frequency bins $\{\omega_i\}$.  The channel output for the
$i$th-frequency mode will then be the coherent state
$|\sum_{k=1}^m \sqrt{\eta_k} \alpha_{k,i}\rangle$, where the
frequency-independent transmissivities $\eta_k$ sum to one. The
input power constraint on the $k$th user is
\begin{equation}
   \sum_i \hbar\omega_i \langle|\alpha_{k,i}|^2\rangle \Delta/2\pi \leq P_k,
\end{equation}
for $1\le k \le m$.  If the receiver uses homodyne or heterodyne
detection, then the wideband capacity region is defined by the
inequalities
\begin{equation}
    \sum_{k\in S} R_k \leq
                            \sqrt{ \sum_{k\in S} \frac{\eta_k P_k}{\pi\hbar}},
\end{equation}
for all $S \subseteq \{1,\ldots, m\}$.

In deriving the wideband capacity region for the two-user,
coherent-state MAC with homodyne or heterodyne detection, we first
obtained upper bounds on the individual rates $R_1$, $R_2$, and
the sum rate $R_1 + R_2$, and then showed that these bounds could
be achieved simultaneously. Applying this same procedure to the
two-user, coherent-state MAC without constraining its receiver
structure, we have obtained the following capacity region,
\begin{subequations}
\label{optwideband}
\begin{align}
   R_1 &\leq \sqrt{\frac{\pi \eta P_A}{3\hbar}}\\
   R_2 &\leq \sqrt{\frac{\pi (1-\eta)P_B}{3\hbar}}\\
   R_1 + R_2 &\leq \sqrt{\frac{\pi [\eta P_A + (1-\eta)P_B]}{3\hbar}},
\end{align}
\end{subequations}
and optimal average photon number allocations,
\begin{align}
  \eta\bar{n}_A(\omega) &= \frac{1}{\exp\!{\left(
                           \sqrt{\pi \hbar \omega^2/12\eta P_A}
                           \right)} - 1}, \\
 (1-\eta) \bar{n}_B(\omega) &= \frac{1}{\exp\!{\left(
                           \sqrt{\pi \hbar \omega^2/12(1-\eta) P_B}
                           \right)} - 1},
\end{align}
\begin{equation}
  \bar{n}'_{AB}(\omega) = \frac{1}{\exp\!{\left(
                           \sqrt{\pi \hbar \omega^2/12[\eta P_A + (1-\eta)P_B]}
                           \right)} - 1}.
\end{equation}
Equations~(\ref{widehom}) and (\ref{optwideband}) show that
optimal reception increases  both the individual rates and the sum
rate by a factor of $\pi/\sqrt{3}$ as compared to what is
achievable with homodyne or heterodyne detection.  The $m$-user
capacity region for the coherent-state MAC is specified by the
inequalities
\begin{equation}
    \sum_{k\in S} R_k \leq
                            \sqrt{ \sum_{k\in S} \frac{\pi\eta_k P_k}{3\hbar}},
\end{equation}
for all $S \subseteq \{1,\ldots, m\}$; once again there is an
improvement factor of $\pi/\sqrt{3}$ as compared to homodyne or
heterodyne detection.

\section{Gaussian MAC}
\label{sec:gaussianstatemac}

Now let us return to the single-mode case and relax our assumption
that the transmitters use coherent-state encodings, i.e., we will
allow them to use non-classical states in their quest for the
largest possible capacity region.  As a step toward finding the
ultimate capacity region of the optical MAC, let us allow Alice
and Bob to employ arbitrary Gaussian states, instead of just
coherent states.

\subsection{Holevo-Sohma-Hirota MAC}
\label{sec:HSHmac}

It is useful to begin by deriving  the capacity region for the
two-user version of the Holevo-Sohma-Hirota \cite{holevo99} (HSH)
channel model. Consider a mode with annihilation operator $\hat{a}
= \hat{a}_1 + i\hat{a}_2$ that is in a zero-mean, Gaussian state,
$\hat{\rho}(0)$, with quadrature-component covariance matrix
\begin{equation}
   V = \begin{pmatrix}
        V_1 & V_{12} \\[.1cm]
        V_{12} & V_2
       \end{pmatrix}
     \equiv
      \begin{pmatrix}
        \langle \hat{a}_1^2 \rangle & \langle \hat{a}_1 \hat{a}_2 + \hat{a}_2 \hat{a}_1 \rangle/2 \\[0.1cm]
         \langle \hat{a}_1 \hat{a}_2 + \hat{a}_2 \hat{a}_1\rangle/2 & \langle \hat{a}_2^2 \rangle
        \end{pmatrix}.
\end{equation}
We define a multiple access channel model in which Alice and Bob
send classical messages $\alpha$ and $\beta$, subject to input
constraints
\begin{align}
  \langle|\alpha|^2\rangle \equiv \int\! |\alpha|^2 p_A(\alpha) \,{\rm d}\alpha & = N_A, \label{macconstraintA}\\
  \langle|\beta|^2\rangle \equiv \int\! |\beta|^2 p_B(\beta) \,{\rm d}\beta & = N_B,
\label{macconstraintB}
\end{align}
and Charlie receives the shifted version of the initial state,
viz., $\hat{\rho}(\alpha,\beta)
  = \hat{D}(\alpha+\beta) \hat{\rho}(0) \hat{D}^{\dagger}(\alpha+\beta)$,
  where
 $\hat{D}(\gamma) \equiv \exp(\gamma\hat{a}^\dagger - \gamma^*\hat{a})$ is the displacement operator.
From the quantum MAC capacity theorem \cite{winter}, the capacity
region of the two-user HSH MAC is given by the convex hull of all
rate pairs $(R_1, R_2)$ satisfying
\begin{subequations}
\begin{align}
  R_1 &\leq S(\bar{\rho}_A) - S(\hat{\rho}(0)) \label{gaussianmac1}\\
  R_2 &\leq S(\bar{\rho}_B) - S(\hat{\rho}(0))\\
  R_1+R_2 &\leq S(\bar{\rho}_{AB})- S(\hat{\rho}(0)),
  \label{gaussianmac12}
\end{align}
\end{subequations}
for some product distribution $p_A(\alpha) p_B(\beta)$, where the
average density operators are
\begin{align}
  \bar{\rho}_A &= \int\! p_A(\alpha) \hat{D}(\alpha)\hat{\rho}(0)\hat{D}^{\dagger}(\alpha) \,{\rm d}\alpha, \\
  \bar{\rho}_B &= \int\! p_B(\beta) \hat{D}(\beta)\hat{\rho}(0)\hat{D}^{\dagger}(\beta) \,{\rm d}\beta, \\
  \bar{\rho}_{AB} &=  \int\!\!\int\! p_A(\alpha)p_B(\beta) \hat{\rho}(\alpha,\beta)
                 \,{\rm d}\alpha \,{\rm d}\beta.
\end{align}
To evaluate this capacity region, we will first maximize the rate
upper bounds for $R_1$, $R_2$, and $R_1 + R_2$ separately. Then we
will then show that the region described by these maximum rates is
achievable.

To maximize the $R_1$ upper bound in \eqref{gaussianmac1}, we
follow the proof of the HSH  capacity theorem \cite{holevo99}; the
same derivation will also apply to the $R_2$ upper bound. For any
input distribution $p_A(\alpha)$ that satisfies constraint
\eqref{macconstraintA}, let $\tilde{p}_A(\alpha)$ be the zero-mean
Gaussian distribution with the same second moments as
$p_A(\alpha)$.  Then, $\tilde{p}_A(\alpha)$ satisfies constraint
\eqref{macconstraintA} and
\begin{equation}
 \tilde{\rho}_A =  \int\! \tilde{p}_A(\alpha)
  \hat{D}(\alpha) \hat{\rho}(0) \hat{D}^{\dagger}(\alpha) \,{\rm d}\alpha
\end{equation}
is a Gaussian state.  If $F(\hat{a},\hat{a}^{\dagger})$ is any
second-order polynomial in $(\hat{a}, \hat{a}^{\dagger})$, then
\begin{align}
     \text{tr}[ \bar{\rho}_A F(\hat{a},\hat{a}^{\dagger})]
  &=  \int\! p_A(\alpha)
                  \,\text{tr}[
                    \hat{D}(\alpha) \hat{\rho}(0) \hat{D}^{\dagger}(\alpha)
                     F(\hat{a},\hat{a}^{\dagger})]\,
                   {\rm d}\alpha  \\
  &=  \int p_A(\alpha)
                  \,\text{tr}[
                    \hat{\rho}(0) F(\hat{a}+\alpha, \hat{a}^{\dagger}+\alpha^*)]\,
                   {\rm d}\alpha  \\
  &=  \int \tilde{p}_A(\alpha)
                  \,\text{tr}[
                    \hat{\rho}(0) F(\hat{a}+\alpha,\hat{a}^{\dagger}+\alpha^*)]\,
                   {\rm d}\alpha \\
  &= \text{tr}[ \tilde{\rho}_A F(\hat{a},\hat{a}^{\dagger})].
\end{align}
Thus, $\bar{\rho}_A$ and $\tilde{\rho}_A$ have the same second
moments, and it follows that $S(\tilde{\rho}_A) \geq
S(\bar{\rho}_A)$, i.e., we can restrict our attention to Gaussian
input distributions in trying to maximize the $R_1$ upper bound.

When the input distribution $p_A(\alpha)$ is Gaussian, the rate
upper bound for $R_1$ can be expressed as
\begin{equation}
     S(\bar{\rho}_A) - S(\hat{\rho}(0))
   = g(2|V + V_{\alpha}|^{1/2} - 1/2)
     - g(2|V|^{1/2} - 1/2),
\end{equation}
where the quadrature-component covariance matrix of $p_A(\alpha)$
is
\begin{equation}
   V_{\alpha}
    = \begin{pmatrix}
        V_1^{\alpha} & V_{12}^{\alpha} \\[.1cm]
        V_{12}^{\alpha} & V_2^{\alpha}
       \end{pmatrix}.
\end{equation}

Thus, the optimization problem we need to solve is
 $\max_{V_{\alpha}} |V + V_{\alpha}|$,
subject to the positive semidefinite and input power constraints
\begin{align}
   V_{\alpha} &\geq 0, \\
   \text{tr}(V_{\alpha}) &= V_1^{\alpha} + V_2^{\alpha} = N_A.
\end{align}
This constraint region is the interior of a circle in the
$V_1^{\alpha} - V_{12}^{\alpha}$ plane, which has the following
polar-coordinate parameterization,
\begin{equation}
  V_1^{\alpha} = r \cos\theta + \frac{N_A}{2},\quad
  V_{12}^{\alpha} = r\sin\theta,\quad
  V_2^{\alpha} = -r\cos\theta + \frac{N_A}{2},
\end{equation}
where $0\leq r\leq N_A/2$ and $0\leq\theta < 2\pi$.  Now, write
\begin{align}
     |V + V_{\alpha}|
     &= (V_1 + V_1^{\alpha})(V_2 + V_2^{\alpha}) - (V_{12} + V_{12}^{\alpha})^2 \\
     &= \left(\frac{V_1 + V_2 + N_A}{2}\right)^2
        - \left(\frac{V_1 - V_2}{2} + r\cos\theta \right)^2
        \nonumber\\
        &\qquad- (V_{12} + r\sin\theta)^2.
\end{align}
In terms of $r$ and $\theta$, our maximization problem then
becomes
\begin{align}
  \max_{V_{\alpha}} &  |V + V_{\alpha}|  \nonumber\\
      &=  \left(\frac{V_1 + V_2 + N_A}{2}\right)^2
         \nonumber\\
      & -\min_{r, \theta}
          \left[
            \left(\frac{V_2 - V_1}{2} - r\cos\theta \right)^2
              + (-V_{12} - r\sin\theta)^2
          \right].
\label{maximizationparameter}
\end{align}
This maximization has two different solutions, depending on
whether or not the point $((V_2 - V_1)/2,$\linebreak $-V_{12})$
lies in the radius-$N_A/2$ circle whose center is at the origin.
If $((V_2 - V_1)/2,-V_{12})$ lies in this circle, then the minimum
on the right-hand side of \eqref{maximizationparameter} is zero.
If \mbox{$((V_2 - V_1)/2, -V_{12})$} lies outside this circle,
then a simple geometric calculation gives the minimum on the
right-hand side of \eqref{maximizationparameter}. We thus obtain
the maximum individual rates
\begin{widetext}
\begin{eqnarray}
    \lefteqn{R_{\text{max}1}
     = \max_{V_{\alpha}}
         S(\bar{\rho}_A) - S(\hat{\rho}(0))} \\
     &=&
     \left\{
       \begin{array}{l}
       g\!\left(V_1 + V_2 + N_A - \frac{1}{2}  \right)
        - g\left(2|V|^{1/2} - \frac{1}{2}\right),\\[.05in]
        \hspace{2cm}
        \text{for $N_A \geq ((V_1 - V_2)^2 + 4V_{12}^2)^{1/2}$}, \\[.5cm]
      g\!\left(2\! \left[
         [(V_1 + V_2 + N_A)/2]^2 -
            \left(
             \sqrt{[(V_1 - V_2)/2]^2 +
                    V^{2}_{12}} - N_A/2
            \right)^2
            \right]^{1/2} - \frac{1}{2}\right)
         - g\left(2|V|^{1/2} - \frac{1}{2}\right),\\[.05in]
      \hspace{2cm}
      \text{for $N_A < ((V_1 - V_2)^2 + 4V_{12}^2)^{1/2}$}
      \end{array}
       \right.
\end{eqnarray}
\end{widetext}
A similar expression holds for the maximum rate $R_{\text{max}2}$.
The fact that capacity is given by two different expressions
depending on whether input constraints satisfy a certain
inequality is referred to as a noncommutative generalization of
waterfilling in \cite{holevo99}.

To maximize the  sum-rate upper bound, we follow the same
approach. It is again sufficient to consider Gaussian input
distributions $p_A(\alpha)$ and $p_B(\beta)$, so our maximization
problem is $\max_{V_{\alpha}, V_{\beta}} |V + V_{\alpha} +
V_{\beta}|$, subject to the positive semidefinite and input power
constraints
\begin{align}
   V_{\alpha} &\geq 0,\quad
   V_{\beta} \geq 0, \\
   \text{tr}(V_{\alpha}) &= V_1^{\alpha} + V_2^{\alpha} = N_A,\quad
   \text{tr}(V_{\beta}) = V_1^{\beta} + V_2^{\beta} = N_B.
\end{align}
This constraint region is the interior of two circles whose
polar-coordinate parameterizations are
\begin{equation}
  V_{\alpha} = \begin{pmatrix}
                   r_A\cos\theta_A & r_A\sin\theta_A \\
                   r_A\sin\theta_A & -r_A\cos\theta_A
                   \end{pmatrix}
                   + \frac{N_A}{2}I,
\end{equation}
\begin{equation}
  V_{\beta} = \begin{pmatrix}
                   r_B\cos\theta_B & r_B\sin\theta_B \\
                   r_B\sin\theta_B & -r_B\cos\theta_B
                   \end{pmatrix}
                   + \frac{N_B}{2}I,
\end{equation}
for $0\leq r_A \leq N_A/2$, $0\leq r_B\leq N_B/2$, $0\leq\theta_A
< 2\pi$, and $0\leq\theta_B < 2\pi$, with $I$ being the $2\times
2$ identity matrix. This parameterization allows us to write
\begin{eqnarray}
 |V + V_{\alpha} + V_{\beta}|
      &=&   (V_1 + V_1^{\alpha} + V_1^{\beta})(V_2 + V_2^{\alpha} + V_2^{\beta}) \nonumber\\
         && - (V_{12} + V_{12}^{\alpha} + V_{12}^{\beta})^2  \\
      &=&  \left(\frac{V_1 + V_2 + N_A + N_B}{2}\right)^2 \nonumber\\
         && - \left(\frac{V_1 - V_2}{2} + r_A\cos\theta_A + r_B\cos\theta_B \right)^2 \nonumber\\
         && - (V_{12} + r_A\sin\theta_A + r_B\sin\theta_B)^2.
\end{eqnarray}
Thus, we have
\begin{align}
  \max_{V_{\alpha}, V_{\beta}}  &   |V + V_{\alpha} + V_{\beta}| \nonumber\\
    &= \left(\frac{V_1 + V_2 + N_A + N_B}{2}\right)^2  \nonumber\\
        &  - \min_{r_A, r_B, \theta_A, \theta_B}
         \left[
            \left(\frac{V_2 - V_1}{2} - r_A\cos\theta_A - r_B\cos\theta_B\right)^2
           \right. \nonumber\\
        &\qquad\qquad\qquad   \Biggl.
              + (-V_{12} - r_A\sin\theta_A - r_B\sin\theta_B)^2
          \Biggr].
\label{maximizationparameterAB}
\end{align}
The second term on the right in \eqref{maximizationparameterAB} is
the minimum squared distance between the points $((V_2 - V_1)/2,
-V_{12})$ and $(r_A\cos\theta_A + r_B\cos\theta_B, \;
r_A\sin\theta_A + r_B\sin\theta_B)$. If $((V_2 - V_1)/2, -V_{12})$
lies in the radius-$(N_A + N_B)/2$ circle that is centered at the
origin, then the second term vanishes. Otherwise, a simple
calculation gives this minimum distance. As a result we find that
the maximum sum rate is,
\begin{widetext}
\begin{eqnarray}
    \lefteqn{R_{\text{max}12}
     = \max_{V_{\alpha}, V_{\beta}}
         S(\bar{\rho}_{AB}) - S(\hat{\rho}(0))} \\
     &=&
     \left\{
       \begin{array}{l}
       g\!\left(V_1 + V_2 + N_A + N_B - \frac{1}{2}  \right)
        - g\!\left(2|V|^{1/2} - \frac{1}{2}\right),\\[.05in]
        \hspace{2cm}
        \text{for $N_A + N_B \geq ((V_1 - V_2)^2 + 4V_{12}^2)^{1/2}$} \\[.5cm]
      g\!\left(2\! \left[
            [
             (V_1 + V_2 + N_A + N_B)/2
            ]^2 -
            \left(
             \sqrt{[(V_1 - V_2)/2]^2 +
                    V^{2}_{12}} - (N_A + N_B)/2
            \right)^2
            \right]^{1/2} - \frac{1}{2}
            \right)
         \\
      \qquad - g\!\left(2|V|^{1/2} - \frac{1}{2}\right),\\[.05in]
      \hspace{2cm}
      \text{for $N_A + N_B < ((V_1 - V_2)^2 + 4V_{12}^2)^{1/2}$.}
      \end{array}
       \right.
\end{eqnarray}
\end{widetext}

We claim that the two-user capacity region for the HSH MAC with
initial-state quadrature-component variance matrix $V$ and input
constraints $N_A$ and $N_B$, is the region defined by the
inequalities
\begin{equation}
  R_1 \leq R_{\text{max}1},\quad
  R_2 \leq R_{\text{max}2},\quad\mbox{and}\quad
  R_{12} \leq R_{\text{max}12}.
\end{equation}
To verify this claim, we will show that the corners of this region
are achievable.  The capacity region result then follows by
time-sharing.

\begin{figure}
\centering
\includegraphics[width=5.5cm, height=5.5cm]{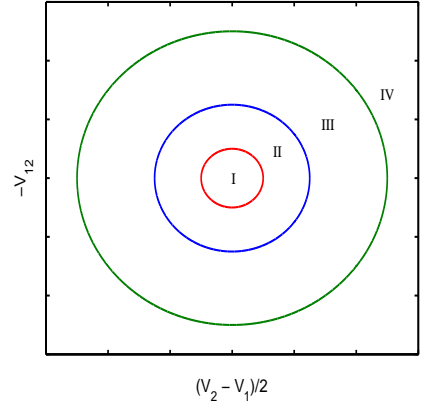}
\caption{\small (Color online) Regions used to complete the HSH
MAC capacity-region proof.  Case I: $r_V \leq N_A/2$. Case II:
$N_A/2 < r_V \leq N_B/2$. Case III: $N_B/2 < r_V \leq (N_A +
N_B)/2$. Case IV: $r_V > (N_A + N_B)/2.$} \label{fig:regions}
\end{figure}

Let the point $((V_2 - V_1)/2, -V_{12})$ have coordinates $(r_V,
\theta_V)$ and suppose that $N_B > N_A$.  To show that the lower
corner $(R_{\text{max}1}, R_{\text{max}12} - R_{\text{max}2})$ is
achievable, we need to find points $(r_A, \theta_A)$ and $(r_B,
\theta_B)$ that simultaneously minimize the distance between
$(r_V, \theta_V)$ and $(r_A, \theta_A)$ and the distance between
$(r_V, \theta_V)$ and $(r_A, \theta_A) + (r_B, \theta_B)$.
Similarly, to show that the upper corner $(R_{\text{max}12} -
R_{\text{max}1},R_{\text{max}2} )$ is achievable, we need to
minimize the distance between $(r_V, \theta_V)$ and $(r_B,
\theta_B)$ and the distance between $(r_V, \theta_V)$ and $(r_A,
\theta_A) + (r_B, \theta_B)$.  There are four cases to consider:
see Fig.~\ref{fig:regions}.  For each case, we list the
coordinates $(r_A, \theta_A)$ and $(r_B, \theta_B)$ corresponding
to the capacity-achieving input distributions.
\begin{itemize}
\item Case I.
  \begin{itemize}
  \item lower corner: $(r_A, \theta_A) = (r_V, \theta_V)$ and $(r_B, \theta_B) = (0,0)$
  \item upper corner: $(r_A, \theta_A) = (0,0)$ and $(r_B, \theta_B) = (r_V, \theta_V)$
  \end{itemize}
\item Case II.
  \begin{itemize}
  \item lower corner: $(r_A, \theta_A) = (N_A/2, \theta_V)$ and $(r_B, \theta_B) = (r_V - N_A/2,\theta_V)$
  \item upper corner: $(r_A, \theta_A) = (0,0)$ and $(r_B, \theta_B) = (r_V, \theta_V)$
  \end{itemize}
\item Case III.
\begin{itemize}
  \item lower corner: $(r_A, \theta_A) = (N_A/2, \theta_V)$ and $(r_B, \theta_B) = (r_V-N_A/2,\theta_V)$
  \item upper corner: $(r_A, \theta_A) = (r_V - N_B/2,\theta_V)$ and $(r_B, \theta_B) = (N_B/2, \theta_V)$
  \end{itemize}
\item Case IV.
\begin{itemize}
  \item lower corner: $(r_A, \theta_A) = (N_A/2, \theta_V)$ and $(r_B, \theta_B) = (N_B/2, \theta_V)$
  \item upper corner: $(r_A, \theta_A) = (N_A/2, \theta_V)$ and $(r_B, \theta_B) = (N_B/2, \theta_V)$.
  \end{itemize}
\end{itemize}

\subsection{Gaussian MAC Capacity}

We now apply the capacity result derived in the previous section
to the two-user optical MAC in Fig.~\ref{mac}. Alice and Bob
encode their classical messages $\alpha$ and $\beta$ using input
states of the form
\begin{align}
  \hat{\rho}_A(\alpha) &= \hat{D}(\alpha)\hat{\rho}_A(0)\hat{D}^{\dagger}(\alpha) \\
  \hat{\rho}_B(\beta)  &= \hat{D}(\beta)\hat{\rho}_B(0)\hat{D}^{\dagger}(\beta),
\end{align}
where $\hat{\rho}_A(0)$ and $\hat{\rho}_B(0)$ are zero-mean
Gaussian states with quadrature-component covariance matrices
$V_A$ and $V_B$, respectively.  This is a modulation code for
which the coherent-state encoding is the special case in which
$\hat{\rho}_A(0)$ and $\hat{\rho}_B(0)$ are vacuum states.
Charlie receives the output ensemble $ \{ p_A(\alpha)p_B(\beta),
\;
      \mathcal{E}(\hat{\rho}_A(\alpha)\otimes\hat{\rho}_B(\beta)) \}$,
where the channel output
$\mathcal{E}(\hat{\rho}_A(\alpha)\otimes\hat{\rho}_B(\beta))$ is
the Gaussian state with mean $\sqrt{\eta}\,\alpha +
\sqrt{1-\eta}\,\beta$ and covariance matrix $\eta V_A + (1-\eta)
V_B$. The capacity of this Gaussian MAC, with input mean photon
number constraints $\bar{n}_A$ and $\bar{n}_B$, is the HSH
capacity region found in the previous section with
\begin{align}
    V &= \eta V_A + (1-\eta)V_B, \\
    N_A &= \eta\!\left(\bar{n}_A - V_1^A - V_2^A + 1/2\right), \\
    N_B &= (1-\eta)\!\left(\bar{n}_B - V_1^B - V_2^B + 1/2\right).
\end{align}
For $\bar{n}_A$ and $\bar{n}_B$ sufficiently large, this capacity
region is the set of rate pairs that satisfy
\begin{subequations}
\label{gaussianmacregion}
\begin{eqnarray}
   R_1 &&\leq  g\!\left(\eta\bar{n}_A
          + (1-\eta)\left(V_1^B + V_2^B - 1/2\right)\right) \nonumber\\
        &&\qquad - g\!\left(2|V|^{1/2} - 1/2\right),
             \label{gaussian1} \\[.1cm]
   R_2 &&\leq  g\!\left(\eta\left(V_1^A + V_2^A - 1/2\right) + (1-\eta)\bar{n}_B\right) \nonumber\\
        &&\qquad - g\!\left(2|V|^{1/2} - 1/2\right),
\end{eqnarray}
\begin{equation}
   R_1 + R_2 \leq g(\eta \bar{n}_A + (1-\eta)\bar{n}_B)
                 - g\!\left(2|V|^{1/2} - 1/2\right).\label{gaussian12}
\end{equation}
\end{subequations}
When $V_A = V_B = I/4$, Alice's and Bob's initial states are
vacuum states, hence they are employing coherent-state encoding
and Eqs.~\eqref{gaussianmacregion} reduce to the coherent-state
formulas in (\ref{coherentstatemac}).
  As shown in
Fig.~\ref{fig:gaussianmac}, it is possible to find $V_A$ and
$V_B$, for example,
\begin{equation}
V_A = V_B =
  \begin{pmatrix}
    1/32 & 0 \\[.05in]
     0 & 2
  \end{pmatrix},
\label{inputvariances}
\end{equation}
when $\eta = 1/2$, $\bar{n}_A = 10$, and $\bar{n}_B = 8$, such
that the Gaussian MAC capacity region is larger than the
coherent-state MAC region.  Numerical search over the space of
possible covariance matrices is one way to further enlarge the
capacity region beyond that achieved by this example.
In the next
section, we derive a result that implies the maximum individual
rates achievable over the Gaussian MAC when Alice, say, is
allowed to choose the optimal input covariance matrix $V_A$ corresponding
to Bob's covariance matrix $V_B$.
In Section~\ref{sec:outerbounds}, we show that transmitting Gaussian
states is asymptotically optimal in the limit of large $\bar{n}_A$
and $\bar{n}_B$.

\section{Anisotropic Gaussian-Noise Capacity}

In this section, we generalize previous work
\cite{lossychannel,freespace,qcmc04,outputentropy} on single-user
lossy Bosonic channels with Gaussian excess noise to include
anisotropic (colored) noise.   In this section, the channel model
we shall consider is the trace-preserving completely-positive
(TPCP) map, $\mathcal{E}^{V_b}_{\eta}(\cdot)$, associated with the
evolution from input mode $\hat{a}$ to output mode $\hat{c} =
\sqrt{\eta}\,\hat{a} + \sqrt{1-\eta}\,\hat{b}$, when the noise
mode, $\hat{b}$, is in a zero-mean Gaussian state, $\hat{\rho}_b$,
with quadrature covariance matrix $V_b$.  Let $\bar{n}_b$ denote
the mean photon number of the Gaussian noise state $\hat{\rho}_b$.

In seeking the capacity of this channel, we shall \emph{assume}
that the conjecture about the minimum output entropy of the
thermal-noise (isotropic-Gaussian noise) channel
\cite{outputentropy} is correct.  This conjecture states that the
minimum output entropy of the thermal-noise channel
$\mathcal{E}(\cdot)$, which has $V_b = (2\bar{n}_T + 1)I/4$, where
$I$ is the $2\times 2$ identity matrix, is given by
\begin{equation}
    \min_{\hat{\rho}} S(\mathcal{E}(\hat{\rho})) = g((1-\eta)\bar{n}_T)
\end{equation}
Presuming the correctness of this conjecture, we now have the
following theorem.
\begin{theorem}
\label{thm:gaussiancapacity} The classical capacity of the
Gaussian-noise channel $\mathcal{E}^{V_b}_{\eta}$ is given by
\begin{equation}
  C = g(\eta\bar{n} + (1-\eta)\bar{n}_b)
    - g\left((1-\eta)\left(2|V_b|^{1/2} - 1/2 \right)\right),
\label{thm:gaussian}
\end{equation}
for input mean photon numbers $\bar{n} \geq
\bar{n}_{\emph{thresh}}$, where
\begin{align}
   \bar{n}_{\emph{thresh}} &= \frac{1}{\eta}
                       \left( (V'_1 - V'_2)^2
                            + 4 V_{12}^{' 2}
                       \right)^{1/2}
                       + V_1 + V_2 - \frac{1}{2}, \\
  V' &= \eta V + (1-\eta) V_b, \\
  V &= \frac{1}{4}\begin{pmatrix}
                 |\mu + \nu|^2 & 2\, {\rm Im}(\mu\nu) \\[.2cm]
                 2\,\emph{Im}(\mu\nu) & |\mu - \nu|^2
                  \end{pmatrix},
\label{squeezevariance}
\end{align}
and the parameters $\mu$ and $\nu$ are chosen such that the
squeeze operator $\hat{S}(z)$ whitens the Gaussian state
$\hat{\rho}_b$.
\end{theorem}
For sufficiently large input mean photon number $\bar{n}$,
\eqref{thm:gaussian} gives the classical capacity of the
single-user Gaussian-noise channel.

\begin{figure}
\centering
\includegraphics[width=8cm]{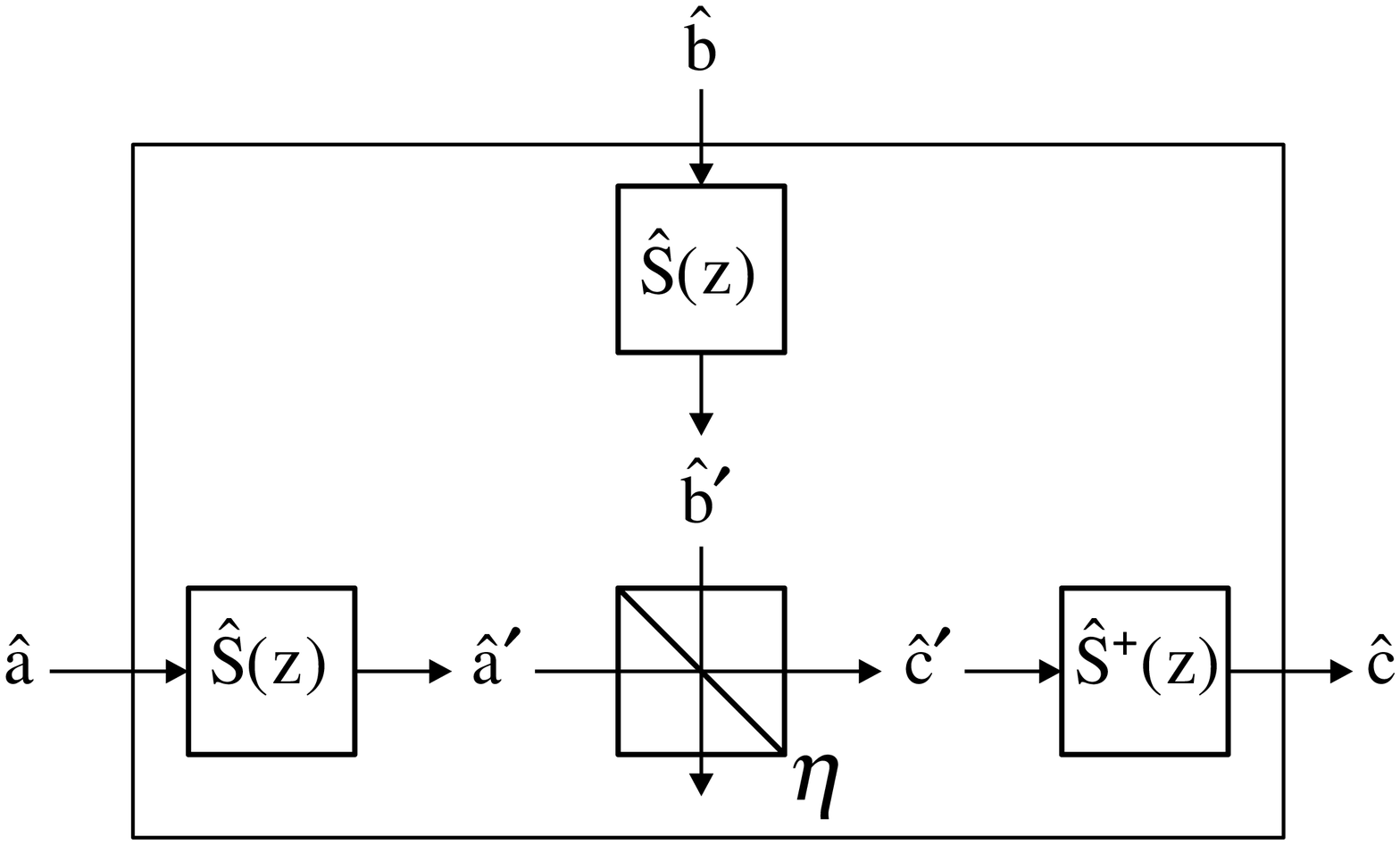}
\caption{\small Equivalent thermal-noise channel
$\mathcal{E}^{\bar{n}_T}_{\eta}$ from $\hat{a}'$ to $\hat{c}'$.
The input mode $\hat{a}'$ is in state $\hat{\rho}'$, and the noise
operator $\hat{b}'$ is in a thermal state with mean photon number
$\bar{n}_T = 2|V_b|^{1/2} - 1/2$. The original Gaussian-noise
channel takes input $\hat{a}$ to output $\hat{c}$.  For the
squeeze operator $\hat{S}(z) \equiv \exp[(z^{*}\hat{a}^2 -
z\hat{a}^{\dagger 2})/2]$, we use the parameterization $\mu =
\cosh r$ and $\nu = e^{i\theta}\sinh r$, where $z = re^{i\theta}$.
Squeezed vacuum states are defined as $|0;z\rangle \equiv
\hat{S}(z)|0\rangle$.} \label{fig:whiten}
\end{figure}

\textbf{Proof} We begin by establishing an upper bound on the
capacity. By the HSW theorem,
\begin{align}
C  &\leq \max_{\{p_j, \hat{\rho}_j\}}
        S\left(\sum_j p_j \mathcal{E}^{V_b}_{\eta}(\hat{\rho}_j)\right)
        - \min_{\{p_j, \hat{\rho}_j\}}
        \sum_j p_j S(\mathcal{E}^{V_b}_{\eta}(\hat{\rho}_j)) \\
   &\leq g(\eta\bar{n} + (1-\eta)\bar{n}_b) -
        \min_{\hat{\rho}_j}S(\mathcal{E}^{V_b}_{\eta}(\hat{\rho}_j)).
        \label{gaussianupperbd}
\end{align}
As sketched in Fig.~\ref{fig:whiten}, we can use the unitary
squeeze operator $\hat{S}(z)$ to find a thermal-noise channel,
with TPCP map $\mathcal{E}(\cdot)$, whose output minimum output
entropy is equal to that of our anisotropic noise channel. [In
essence, this is the quantum equivalent of the noise-whitening
approach to communication through colored noise that is employed
in classical communication theory.]  The average noise-photon
number, $\bar{n}_T$, of this equivalent channel is
\begin{equation}
\bar{n}_T = 2|V_b|^{1/2} - 1/2,
\end{equation}
which, when used in conjunction with \eqref{gaussianupperbd} and
our minimum output entropy conjecture, shows that the right-hand
side of \eqref{thm:gaussian} is an upper bound on the channel
capacity.

To show that the right-hand side of \eqref{thm:gaussian} is also a
lower bound on the channel capacity when $\bar{n} \geq
\bar{n}_{\text{thresh}}$, we evaluate the information rate
achieved by a single-use squeezed-state code.  Let $\hat{\rho}^0_a
= |0;-z\rangle\langle 0;-z|$ be the zero-mean squeezed state whose
quadrature-component covariance matrix is given by
\eqref{squeezevariance}. Consider that random code in which we
transmit the displaced squeezed states,
\begin{equation}
\hat{\rho}_a(\alpha) =
\hat{D}(\alpha)\hat{\rho}^0_a\hat{D}^{\dagger}(\alpha),
\end{equation}
that are selected with a zero-mean Gaussian probability density
function whose quadrature-component covariance matrix is denoted
$V_a$. Imposing the average photon number constraint,
$\langle\hat{a}^{\dagger}\hat{a}\rangle \leq \bar{n}$, assuming
that $\bar{n}\geq\bar{n}_{\text{thresh}}$, and applying the HSH
capacity result \cite{holevo99}, we find that there is a
squeezed-state code whose information rate equals the right-hand
side of (36).  This implies that
\begin{equation}
 C \geq g(\eta\bar{n} + (1-\eta)\bar{n}_b)
    - g\left((1-\eta)\left(2|V_b|^{1/2} - 1/2 \right)\right).
\end{equation}
Equations (41) and (44) provide coincident upper and lower bounds
on the capacity, when $\bar{n}\geq\bar{n}_{\text{thresh}}$, hence
the proof is complete.

There are two special cases of this theorem that are worth
discussing.  First, it is easy to see that when
\begin{equation}
   V_b = \frac{2\bar{n}_b + 1}{4}I,
\end{equation}
$\mathcal{E}^{V_b}_{\eta}$ reduces to the thermal-noise channel
$\mathcal{E}$.  Theorem 1 then predicts $\bar{n}_{\text{thresh}} =
0$ and $C = g(\eta\bar{n} + (1-\eta)\bar{n}_T) -
g((1-\eta)\bar{n}_T)$, in accord with the capacity conjecture for
the thermal-noise channel \cite{outputentropy}.  A more
interesting special case occurs when $\hat{\rho}_b =
|0;z\rangle\langle 0;z|$ is a squeezed state, with $|\nu|>0$,
i.e., a pure-state anisotropic Gaussian noise.  Here we find
\begin{equation}
  V' = V = V_b = \frac{1}{4}\begin{pmatrix}
                 |\mu - \nu|^2 & -2\, {\rm Im}(\mu\nu) \\[.2cm]
                 -2\, {\rm Im}(\mu\nu) & |\mu + \nu|^2
                  \end{pmatrix},
\end{equation}
which yields
\begin{equation}
  C = g(\eta\bar{n} + (1-\eta)|\nu|^2),
\end{equation}
for $\bar{n}\geq\bar{n}_{\text{thresh}} = |\mu\nu|/\eta +
|\nu|^2$. Note that this capacity is \emph{higher} than that of
the thermal-noise channel with the same $\bar{n}_b$ value.  In
other words, phase-sensitive, pure-state Gaussian noise enhances,
rather than degrades channel capacity for
$\bar{n}\geq\bar{n}_{\text{thresh}}$.

\section{Capacity Outer Bound}
\label{sec:outerbounds}

Achieving the ultimate capacity region of the optical MAC may
require the use of non-Gaussian states, so the capacity of the
Gaussian MAC is still only an inner bound on this region. In this
section, we develop an outer bound on the ultimate capacity region
of the optical MAC.  Let Alice and Bob use input states---averaged
over their respective random-coding ensembles---$\bar{\rho}_A$ and
$\bar{\rho}_B$ that are subject to the average photon number
constraints $\bar{n}_A$ and $\bar{n}_B$. Because von Neumann
entropy is invariant to mean fields, we know that the optimum
$\bar{\rho}_A$ and $\bar{\rho}_B$ will be zero-mean-field states.
This, in turn, implies that $\langle \hat{c}^\dagger\hat{c}\rangle
= \eta\bar{n}_A + (1-\eta)\bar{n}_B$, from which it is easily
shown that
\begin{equation}
  R_1 + R_2 \leq S(\mathcal{E}(\bar{\rho}_A \otimes \bar{\rho}_B)) \\
            \leq g\left( \eta \bar{n}_A + (1-\eta) \bar{n}_B
                 \right).
\label{outer12}
\end{equation}
The sum-rate upper bound in \eqref{outer12} coincides with the
coherent-state MAC result appearing in \eqref{coherentstatemac12}.
Hence, we have shown that the sum rate for the capacity region is
achieved by coherent-state encoding in conjunction with optimum
(joint-measurement) reception.  More generally, the Gaussian-state
encoding is a sum-rate-achieving code in the above-threshold
regime, i.e., \eqref{gaussian12} coincides with \eqref{outer12},
whenever $\mathcal{E}(\rho_A(0)\otimes\rho_B(0))$ is pure.
Moreover, from (\ref{hetref}) it can be shown that heterodyne
reception is asymptotically optimum for the sum rate in the limit
$\eta\bar{n}_A + (1-\eta)\bar{n}_B \rightarrow \infty$.

\begin{figure}
\centering
\includegraphics[width=7cm]{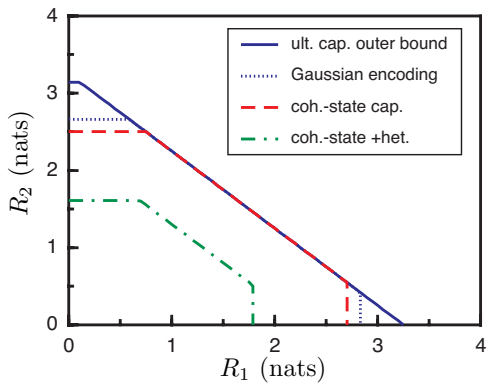}
\caption{\small (Color online) Ultimate capacity region of the
two-user, single-mode optical MAC.  Inner bounds from
coherent-state and Gaussian-state encodings, and outer bounds
given by \mbox{$R_1 \leq g(\bar{n}_A)$}, $R_2 \leq g(\bar{n}_B),$
and $R_1+R_2\leq g(\eta\bar{n}_A + (1-\eta)\bar{n}_B)$ are shown.
The Gaussian-state capacity region is evaluated with input
variance matrices $V_A$ and $V_B$ given by \eqref{inputvariances}.
This figure assumes $\eta = 1/2$, $\bar{n}_A = 10$, and $\bar{n}_B
= 8$.  Rates are measured in nats, i.e., logarithms are taken base
$e$.} \label{fig:gaussianmac}
\end{figure}

To upper bound the individual rates $R_1$ and $R_2$, consider a
super receiver that has access to both output ports of the beam
splitter representing the optical MAC. This super receiver can
invert the unitary beam splitter transformation to undo the
effects of the optical MAC.  Thus, the individual rate upper
bounds reduce to single-user Holevo informations, and we have the
upper bounds $R_1 \leq g(\bar{n}_A)$ and $R_2 \leq g(\bar{n}_B)$.
Our optical MAC results are illustrated in
Fig.~\ref{fig:gaussianmac}.  Here we have plotted the sum rate for
a two-user, single-mode, quantum optical MAC with $\eta = 1/2$,
$\bar{n}_A = 10$, and $\bar{n}_B = 8$, along with the capacity
region for heterodyne detection, the individual rate limits for
coherent-state encoding, and the individual rate limits for the
Gaussian-state encoding from Eq.~\eqref{inputvariances}.

We have presented codes which achieve the sum-rate upper bound,
but it is unknown exactly how far we can reach into the corners of
the outer bound region. One thing we can demonstrate is that the
individual rate upper bounds are asymptotically achievable in the
limit of large $\bar{n}_A$ and $\bar{n}_B$.  Let Alice transmit
real-valued classical messages $\alpha_1$ using squeezed states
$|\alpha_1; z\rangle$ excited in the first quadrature with squeeze
parameter $z>0$. Let Bob transmit the zero-mean squeezed state
$|0; Z\rangle$ with squeeze parameter $Z =
\sinh^{-1}(\sqrt{\bar{n}_B})$, i.e., Bob squeezes as hard as
possible, given his average photon number constraint.   A rate of
\begin{equation}
   R_1 = \frac{1}{2}
         \log\!\left(
         1 + \frac{4(\bar{n}_A - \sinh^2 z)}
                  {e^{-2z} + (1-\eta)e^{-2Z}/\eta}
             \right)
\end{equation}
is achieved if Charlie uses homodyne detection to decode Alice's
message.  After substituting in Alice's optimal value for her
squeeze parameter, $z = \log(2\bar{n}_A + 1)/2$, and performing
several applications of L'H\^{o}pital's rule, we obtain the ratio
\begin{align}
      \lim_{\bar{n}_A \rightarrow\infty} &\lim_{\bar{n}_B \rightarrow\infty}
        \frac{R_1}{g(\bar{n}_A)} \nonumber\\
  &=  \lim_{\bar{n}_A \rightarrow\infty}
        \frac{\frac{1}{2}\log\left(1 + 4e^{2z}(\bar{n}_A - \sinh^2 z)\right)}
             {g(\bar{n}_A)} \\
  &=  \lim_{\bar{n}_A \rightarrow\infty}
        \frac{\log(1 + 2\bar{n}_A)}{g(\bar{n}_A)}\\
  &=  1.
\end{align}
Thus, this squeezed-state code with homodyne detection is
asymptotically optimal for large input photon numbers $\bar{n}_A$
and $\bar{n}_B$.  For the special case $\eta = 1$, Bob is
irrelevant, and the above argument shows that the
squeezed-state/homodyne code is asymptotically optimal for the
single-user lossless Bosonic channel.

\section{Conclusions}
We have derived the capacity region of the Bosonic multiple-access
channel that uses coherent-state encoding.  Single-mode and
wideband transmitters were considered, and in both cases optimum
(joint measurements over entire codewords) reception was shown to
outperform receivers that employed homodyne or heterodyne
detection.  Coherent-state encoding with optimum reception was
shown to achieve the sum-rate bound on the ultimate capacity
region of the optical MAC.  In the limit of high average photon
numbers, the ultimate single-user rates can be achieved with
squeezed-state encoding and homodyne detection.

\begin{acknowledgments}
This work was supported by the DoD Multidisciplinary University
Research Initiative (MURI) program administered by the Army
Research Office under Grant DAAD19--00-1--0177.
\end{acknowledgments}

\appendix*
\section{}

We apply the capacity theorem derived in \cite{winter} based on
the following argument.  Suppose that the transmitters used by
Alice and Bob employ states containing no more than $K$ photons,
where
\begin{equation}
K \gg \max \left\{ 1, \bar{n}_A + \bar{n}_B \right\}.
\label{assume}
\end{equation}
As their states may be indexed by complex-valued parameters
$\alpha$ and $\beta$, over which we can do random coding, the
result described by \eqref{quantummac} specifies the achievable
rate region within this restricted finite-dimensional state space.
The right-hand sides of \eqref{quantummac} --- when maximized over
product distributions that respect the $\bar{n}_A$ and $\bar{n}_B$
constraints --- are monotonically expanding achievable rate
regions with increasing $K$.  Moreover, the achievable rate region
for any $K$ is outer bounded by the results we derive in
Section~\ref{sec:outerbounds} assuming the full Hilbert space is
employed.  For fixed $\bar{n}_A$ and $\bar{n}_B$, the impact of
the truncation to no more than $K$ photons will be negligible,
under the condition \eqref{assume}, and will vanish as $K
\rightarrow \infty$.

\bibliography{pra_mac}

\end{document}